\PassOptionsToPackage{table,xcdraw}{xcolor}
\documentclass[sigconf]{acmart}
\usepackage[shortlabels]{enumitem}
\usepackage{booktabs}
\usepackage{pdfcomment}
\usepackage{framed,enumitem} 
\usepackage{scalerel}
\usepackage{multirow}

\AtBeginDocument{%
  \providecommand\BibTeX{{%
    \normalfont B\kern-0.5em{\scshape i\kern-0.25em b}\kern-0.8em\TeX}}}

\setcopyright{rightsretained}

\begin{document}

\copyrightyear{2023} 
\acmYear{2023} 
\acmConference[CHI '23]{Proceedings of the 2023 CHI Conference on Human Factors in Computing Systems}{April 23--28, 2023}{Hamburg, Germany}
\acmBooktitle{Proceedings of the 2023 CHI Conference on Human Factors in Computing Systems (CHI '23), April 23--28, 2023, Hamburg, Germany}
\acmDOI{10.1145/3544548.3580841}
\acmISBN{978-1-4503-9421-5/23/04}


\title{Relatedly: Scaffolding Literature Reviews with Existing Related Work Sections}

 \author{Srishti Palani}
 \authornote{Work completed during a researcher internship at Semantic Scholar Research, Allen Institute for AI.}
 \email{srishti@ucsd.edu}
 \affiliation{%
  \institution{University of California, San Diego}
  \country{USA}
}

\author{Aakanksha Naik}
\email{aakankshan@allenai.org}
\affiliation{%
  \institution{Allen Institute for AI}
  \city{Seattle}
  \state{WA}
  \country{USA}
}

\author{Doug Downey}
\email{dougd@allenai.org}
\affiliation{%
  \institution{Allen Institute for AI}
  \city{Seattle}
  \state{WA}
  \country{USA}
}
\author{Amy X. Zhang}
\email{axz@cs.uw.edu}
\affiliation{%
  \institution{University of Washington}
  \city{Seattle}
  \state{WA}
  \country{USA}
}

\author{Jonathan Bragg}
\email{jbragg@allenai.org}
\affiliation{%
  \institution{Allen Institute for AI}
  \city{Seattle}
  \state{WA}
  \country{USA}
}

\author{Joseph Chee Chang}
\email{josephc@allenai.org}
\affiliation{%
  \institution{Allen Institute for AI}
  \city{Seattle}
  \state{WA}
  \country{USA}
}

\renewcommand{\shortauthors}{}

\begin{abstract}
Scholars who want to research a scientific topic must take time to read, extract meaning, and identify connections across many papers. As scientific literature grows, this becomes increasingly challenging. Meanwhile, authors summarize prior research in papers’ related work sections, though this is scoped to support a single paper. A formative study found that while reading multiple related work paragraphs helps overview a topic, it is hard to navigate overlapping and diverging references and research foci.
In this work, we design a system, \textit{Relatedly}, that scaffolds exploring and reading multiple related work paragraphs on a topic, with features including dynamic re-ranking and highlighting to spotlight unexplored dissimilar information, auto-generated descriptive paragraph headings, and low-lighting of redundant information. From a within-subjects user study (\textit{n=15}), we found that scholars generate more coherent, insightful, and comprehensive topic outlines using \textit{Relatedly} compared to a baseline paper list.
\end{abstract}

\ccsdesc[500]{Human-centered computing~Graphical user interfaces}

\keywords{Literature Review, Scientific Discovery, Exploratory Search, Sensemaking
}

\maketitle
\section{Introduction}
\label{sec:intro}

\begin{figure*}
\includegraphics[width=\textwidth]{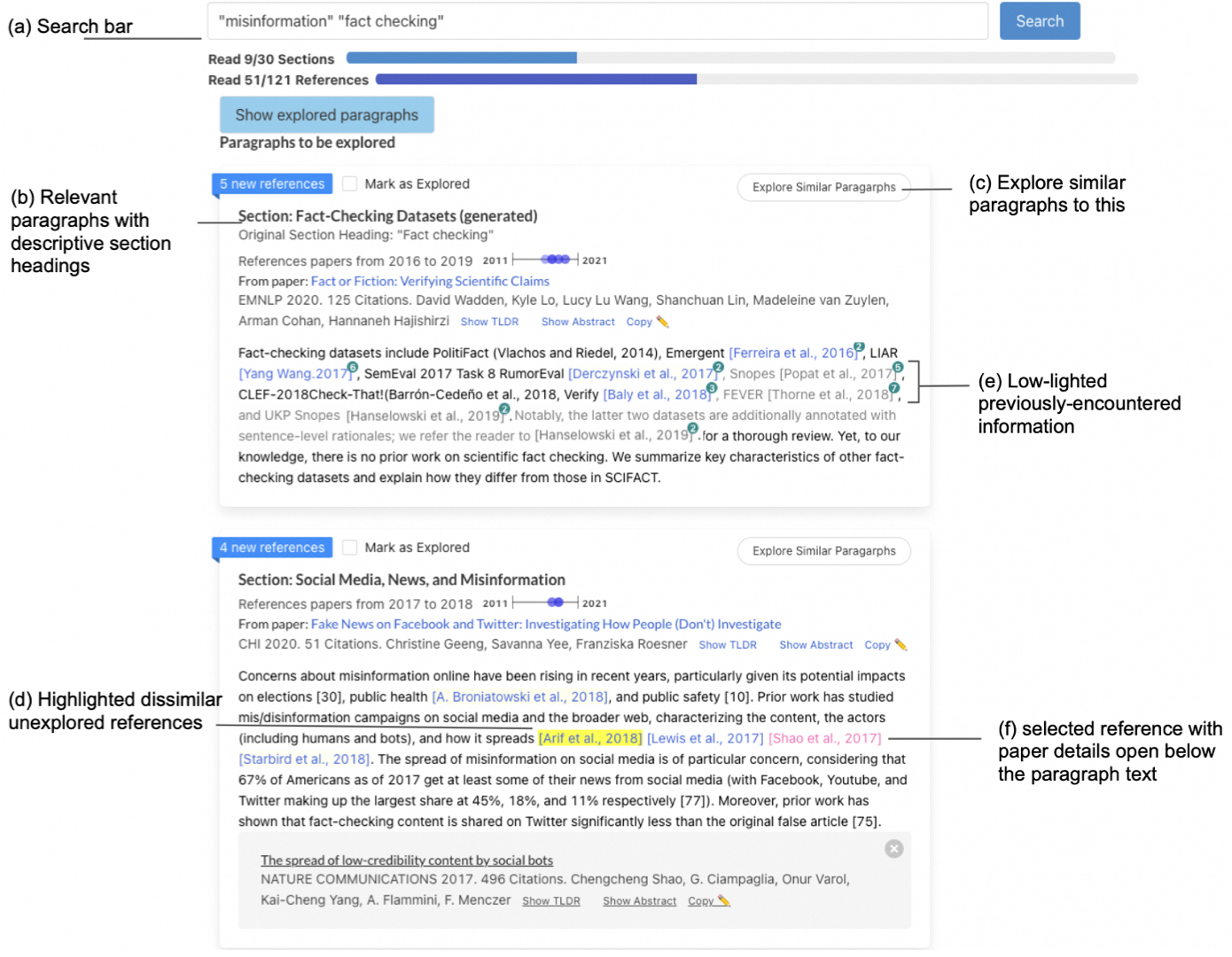}
\caption{The \textit{Relatedly} system presents users with related work paragraphs from prior work on a topic and scaffolds the paragraph exploration experience with features for reading, prioritization, and progress tracking. Here, the \textbf{Overview View} shows paragraphs relevant to the high-level query topic and ranked by diversity so the top results show a wide range of subtopics. A user interested to learn more about one of the paragraph's subtopics could click on the ``Explore Similar Paragraphs'' button, which would take them to the \textbf{Similar Paragraphs View} in Fig~\ref{fig:similarParas}.
}
\label{fig:Relatedly}
\end{figure*} 

Scientific discovery and innovation rely upon scholars to have a rich understanding of prior work, which they achieve through reviewing the literature, extracting meaning, and identifying connections across many papers with large amounts of ambiguous domain-specific information \cite{knopf2006doing, zhang2008citesense}. This process is getting progressively harder with the exponential growth of scientific publications \cite{fortunato2018science, bornmann2021growth, jinha2010article, van2014global} and the increasingly interdisciplinary nature of science \cite{van2015interdisciplinary, okamura2019interdisciplinarity}.
Unfortunately, current approaches such as reading survey papers or using textual or visual search engines are limited in terms of sensemaking support or timeliness. 
For example, survey papers present a broad overview of a research topic with coherent research themes and carefully synthesized descriptions \cite{bruce1994research, knopf2006doing}. But since they require significant manual effort to compile, survey papers are not always available on all topics and can quickly become outdated as new research emerges. To address this, scholars also frequently rely on automatic approaches to help explore literature such as scholarly search engines, including Google Scholar\footnote{\url{scholar.google.com}} and Semantic Scholar\footnote{\url{www.semanticscholar.org}}. These tools can be effective in looking up papers relevant to a query but do not present higher level themes that connect multiple papers.
Other tools use visualization to connect and cluster papers~\cite{king2009scholarly} using metrics based on citations or semantic embedding vectors, such as Connected Papers.\footnote{\url{www.connectedpapers.com}}
However, it can be hard for users to comprehend the underlying meaning of complex graphs and clusters as automatic clusters often conflate multiple dimensions \cite{hearst2006clustering}. As a result, when timely survey papers are not available, scholars still need to examine many individual papers and try to figure out the latent themes and connections between them to conduct literature reviews \cite{pirolli2005sensemaking}.  

Meanwhile,  authors of scholarly papers also go through a similar process of exploring and summarizing prior research whenever they need to write the related work sections of their papers. While a related work section provides up-to-date and well-synthesized summaries of prior work \cite{teevan_2014,tenopir2012article}, because they are scoped to support a single paper they often do not provide a comprehensive overview of the topic like survey papers. However, this issue could potentially be mitigated if readers are presented with \textit{multiple} related work sections about a topic  from different papers so that they gain broader coverage and different perspectives of the space from multiple authors. 

To investigate this opportunity further, we first built a text search engine over a set of related work sections extracted from many papers, and used it to conduct a formative interview study with 10 scholars. We asked scholars about their reactions to and challenges with exploring related work sections when compared to their current practices.
We found that while participants preferred reading related work sections, they had difficulty prioritizing and tracking their reading, given that different related work sections have both overlapping and diverging references and foci.

Motivated by insights from the interviews, we designed \textit{Relatedly}, a novel system for scaffolded exploration of literature that leverages related work sections to provide a synthesis of a broad topic.
As shown in \autoref{fig:Relatedly}, when the user queries a topic in Relatedly (a), the system retrieves relevant paragraphs from different papers' related work sections along with their section headings (b) to help users gain a quick overview of disparate research threads. 
In cases where a paragraph does not have a descriptive section heading, Relatedly automatically generates one.
Users can also drill-down on a subtopic by exploring similar paragraphs for a given paragraph (c, followed by \autoref{fig:similarParas}). 
To support users in prioritizing and tracking their reading, as a user is exploring related work sections in Relatedly, it tracks which paragraphs and references the user has read and then dynamically re-ranks the remaining paragraphs and highlights unexplored references that diverge from the user's history (d).
Relatedly also low-lights sentences that refer to papers that have been cited in already-seen paragraphs (e).

We conducted a within-subjects study ($n=15$) to evaluate Relatedly where participants were asked to explore literature on two scientific topics, with the ultimate goal of producing an outline of a survey paper on each topic using Relatedly in one condition and using a baseline system that returns a list of papers in another.
We find that participants produced better quality outlines when using Relatedly versus in the baseline condition, as rated by topic experts who were blind to the conditions.
System logs reveal that users of Relatedly interacted with significantly more information (both paragraphs and papers) than in the baseline condition, despite having the same amount of time for each condition and access to the same set of papers. Participants also self-reported that they preferred to explore related work sections using Relatedly rather than explore a list of papers to conduct literature review.

In summary, this work makes the following contributions: 

\begin{itemize}
    \item A novel approach to discovering and systematically reviewing literature on a scientific topic by reading and exploring relevant related work sections extracted from many papers. 
    \item Results from a formative user study ($n=10$) outlining current literature review practices and user challenges with this approach. 
    \item The \textit{Relatedly} system, which scaffolds related work paragraph exploration with reading, prioritization, and progress tracking features.
    \item Empirical insights from a within-subjects study with 15 participants 
    that finds that scaffolded exploration of related work sections promotes literature discovery and synthesis. 
\end{itemize}

\section{Related Work}
\label{sec:related}

Our work builds on prior work studying how scientists explore and review literature, and tools built to support these complex exploratory processes. 

\vspace{-2mm} 

\subsection{How Scholars Conduct Literature Reviews} 
Literature review helps scholars identify patterns and gaps in prior research in order to find opportunities, determine rationale for a new investigation, and situate research goals within the literature \cite{teevan_2014}. Reviews detail both known research and open research questions in this topic. A high quality literature review comprehensively includes all the main themes and sub-themes found in a chosen topic of study, from both classic foundational work and recent studies to demonstrate an in-depth understanding of the topic at hand \cite{denney2013write, knopf2006doing, jesson2011doing}. To achieve these goals, scholars must take time to comprehensively explore a topic and read many individual papers. However, the sensemaking process of trying to get an overview of a field from reading individual papers can be time-consuming and cognitively overwhelming 
 \cite{bruce1994research, tenopir2012article, rowland2002overcoming}. For example, it can be hard for users to diversify their readings to quickly identify different threads of research. The overwhelming number of individual papers and redundant information scholars need to go through often leads to information overload \cite{miller1960information}.

One way scholars have addressed this is to write survey papers for different research topics \cite{denney2013write, knopf2006doing, jesson2011doing}. Yet with the exponential increase in publishing rates, survey papers are often unavailable \cite{fortunato2018science, bornmann2021growth, jinha2010article, van2014global}, and even when they are, they quickly get outdated as newer research emerges.  

Meanwhile, in most scientific papers, authors summarize and draw connections across multiple papers to situate their own work in related work sections \cite{tenopir2012article, teevan_2014}. Each paragraph in these related work sections adds context and structure to individual papers referenced. 
For example, the related work section of a paper on misinformation might group a set of referenced papers into a paragraph with a title of ``How misinformation affects public health'', and another set of papers might be grouped under ``How misinformation spreads on social media''. 
However, related work sections only focus on a paper's specific point of view and do not attempt to exhaustively overview all the themes and sub-themes in the broader topic. 
For example, the above paper about misinformation might focus its related work section on health misinformation on social media because that is what is relevant but lack coverage of other work  related to misinformation, such as, say, computational techniques for detecting misinformation. 
Therefore, scholars hoping to gain a broader picture of literature on a topic would likely need to read multiple related work sections across multiple papers. This task is what the Relatedly system is attempting to scaffold.

Information foraging theory \cite{pirolli1995information} provides some pointers on how to go about this task. 
During complex exploratory tasks, people switch between \textit{exploring} different information patches and \textit{exploiting} a discovered patch to optimize information gain. They rely on various cues, or ``information scent'', in the information environment to assess whether a source is promising for gaining information. 
We take inspiration from information foraging theory to provide
information scent cues in Relatedly such as displaying how much new information the user can learn about by reading each paragraph. 
Also, to support switching from exploring to exploiting, Relatedly allows a user to dive in to view similar paragraphs given a paragraph; this enables them to gain a deeper understanding of a sub-topic from different perspectives.

\subsection{Tools for Supporting Literature Review}
One of the most common tools scholars rely on today for literature review is scholarly search engines \cite{tenopir2012article}, such as Google Scholar$^1$ and Semantic Scholar$^2$. These can be very effective in helping users look up individual papers relevant to a query. However, to gain deeper understanding of a research area, such as during literature reviews, scholars often need to synthesize information across individual papers. This effortful and time consuming process of making sense of connections between papers and uncovering the different nuanced research themes within a larger topic is largely left to the users with minimal support \cite{marchionini2006exploratory, russell1993cost}. For example, when exploring papers from a search results list, it can be hard for users to prioritize their readings, keep track of information scattered across multiple papers, or have a sense of their overall progress within the unfamiliar information space.

Faceted search interfaces allow users to navigate search results by applying multiple filters across categories \cite{hearst2006clustering}. Categorizing provides coherent and mostly complete labels. However, manual categorization takes time and effort and is hard to keep updated. Automatic categorization is typically based on metadata \cite{hearst2006clustering}. For example, Google Scholar supports filtering paper results by \textit{time of publication} and \textit{relevance}, among others. Similarly, Semantic Scholar presents \textit{‘fields of study’}, \textit{‘publication types’},  etc., as facets by which papers can be filtered. But metadata is not always available. Also, these labels are often too general and don’t provide meaningful  insight into the topic or domain. 

Visual clustering systems attempt an alternative approach  to help scholars discover relationships between papers. For example, given a seed paper, Connected Papers$^3$ utilizes the citation graph to find clusters of other relevant papers. Research systems like PaperQuest \cite{Ponsard2016PaperQuestAV} and Apolo \cite{chau2011apolo} visualize citation relationships between a set of papers as input, with support to overcome information overload by progressively revealing further related papers given a source paper and its citations. However, prior research in clustering search interfaces has also pointed to how automatically generated clusters can be incoherent and difficult for users to understand because they often conflate multiple dimensions \cite{hearst2006clustering}. Specifically, visual paper clustering approaches often show edges between similar papers but do not describe their semantic relationships \cite{hearst2006clustering}. They also show clusters of similar papers but lack high-level descriptions of the underlying themes \cite{hearst2006clustering}. As a result, scholar still need to examine individual papers to determine the meaning of each automatically generated clusters and how different papers relate to one another \cite{chau2011apolo}.

Automatic summarization techniques like Multi-Document Summarization  \cite{deyoung2021ms2} and Metro Maps of Science \cite{shahaf2012metro} add explanations to otherwise complex and hard to understand citation graphs. However, these explanations are not always accurate, coherent, or comprehensive. On the other hand, manual (e.g., Threddy \cite{kang2022threddy}) and crowd-powered systems (e.g., Knowledge Accelerator \cite{hahn2016knowledge,chang2016alloy, chi23_comlittee}, Crowdlines \cite{luther2015crowdlines}) help provide more coherent, comprehensive, and accurate summaries of topic spaces, but take time and effort to generate. 

Relatedly sidesteps the issue of generating high quality connections and clusters by building on the significant effort that related work authors already expend to construct these for their paper. The main challenge then becomes about exploring multiple papers' overlapping clusters and differing perspectives on how papers connect to each other.

\section{Formative Study \& Design Goals}
\label{sec:formative}
To understand user challenges and strategies when exploring and making sense of related work paragraphs on a topic, we conducted a formative interview study with scholars.

\subsection{Formative User Study Method}
We conducted semi-structured interviews with 10 people who have experience searching for, reading, and writing scientific literature for more than three years (5 male and 5 female, average age of 27.5 years). One had completed their doctoral degree, while five had completed a master's, and four had completed their bachelor's degree. 
In terms of job titles, we had: one post-doctoral researcher, one research assistant, one research scientist and the remaining seven were doctoral researchers. 
Five reported using scholarly web search multiple times a day, four reported doing this at least multiple times a week, and one said rarer than every week. Eight had experience conducting systematic literature reviews for three or more years, one reported doing this for two years, and one for one year. Participants came from diverse domains: neuroscience, geography, biomedical sciences, human-computer interaction, natural language processing, AR/VR design, and wearable computing. They reported mostly using scholarly search engines and paper lists for exploring literature, including Google Scholar, Semantic Scholar, and domain-specific conference proceedings, journals, and organizations (e.g., ACL for NLP or CHI'23 proceedings for HCI). 
 They used a wide range of applications for reading and writing scientific literature.

We asked participants about their workflows for conducting literature review and about any challenges they experience. Then, we gave them 20 minutes to explore and read a set of related work paragraphs extracted from multiple research papers on a topic. The paragraphs were displayed in a list on a simple text search engine interface (see Appendix Fig. \ref{fig:formative}).
To contextualize their exploration, we gave them a simulated task \cite{borlund2003iir} of conducting initial research to get a broad overview of the topic of ``misinformation, fake news and fact-checking'', towards the ultimate goal of writing a literature review. To get insight into their user experience, participants were asked to think-aloud as they explored the list of paragraphs. Afterward, they were asked about their experience reading multiple paragraphs instead of papers, the challenges surrounding this, and strategies they used to overcome these challenges. 
We then presented them with alternative mock-up designs that augment the related work paragraphs with highlighting of  references and terms that are unexplored and low-lighting of redundant citations.

Interviews were conducted remotely over video calls by the first author and lasted around 45 minutes. They were recorded and then transcribed using an auto-transcription service. Then, the first author went through the transcripts and coded them for themes using an open coding approach \cite{charmaz2014constructing}. Through multiple iterations along with periodic discussions with the rest of the research team, we identified the user challenges and subsequently the design goals for our approach. 

\subsection{User Experience when Reviewing Literature}
\subsubsection{Current Literature Review Workflows and Challenges}
When asked about their current workflows, all participants (10/10) mentioned using scholarly search engines to discover relevant papers on a topic. Some (5/10) mentioned socially gathering a list of papers from collaborators, advisors, or social media such as Twitter. All of them mentioned reading  papers one by one to extract meaning, 3/10 mentioned annotating the PDF documents with notes and highlights, and 2/10 mentioned saving these papers to a bibliography manager (e.g., Zotero, Mendeley). 

When asked about challenges, 10/10 mentioned that it was hard to make connections across papers. 8/10 participants mentioned challenges with uknown unknowns ranging from not knowing the right search keywords that would lead to the right papers to not knowing what all the latent subtopics were within the topic of interests: ``\textit{I often don’t know which keywords/domain-specific language to search to get to the right literature''}, \textit{``Even if people refer you to a shortlist of papers, it’s hard to get an overview of the topic and it feels like I might be being myopic and might have blind spots''.} 7/10 mentioned that it was hard to keep track of what they had read before: \textit{``hard to keep track of many different research threads, points of view and see the bigger picture.''} 5/10 discussed challenges prioritizing what to read first : \textit{``When I see so many papers in results, I get overwhelmed and open them up in tabs. But then I don't know which to read first so they will just stay open in these tabs''}. 

\subsubsection{Preference for Exploring Related Work Sections and Challenges}
When asked about their experience reading multiple paragraphs to get a topic overview, 9/10 participants preferred reading related work paragraphs to papers. Some positive reactions discussed getting a broad overview of the topic: \textit{``Reading even a few paragraphs equips me quickly with the relevant vocabulary, references and takeaways from the topic'', ``I’m able to see the different threads of research immediately''}.
Others talked about the value of the text summarizing the referenced papers: \textit{``I like the the additional explanation around the references, so I can understand the context and decide quickly whether I want to open it up to read more or not''}.
Some also indirectly referenced how they already use papers to help find other papers to read, and how extracting related work sections and their citations streamlines the process:
\textit{``Definitely more helpful than reading PDFs and doing the ritual of opening PDFs, reading introductions, and the paper, going back and forth between references in the bibliography and the paper to identify which papers might be useful''}. 

However, there were also challenges with reading multiple paragraphs to get a topic overview.
Some participants desired prioritization and navigation support to know what to read next:
\textit{``hard to prioritize which order to read these in''} (5/10), 
\textit{``it is unclear what the similarities and differences between paragraphs are''} (7/10),
\textit{``want to know which are the most important or central papers summarized in this paragraph''} (3/10).
Participants also wanted support for tracking their exploration: \textit{``want to keep track of what [paper and paragraph] has been read vs not''} (2/10),  \textit{``hard to assess how much more there is to read on this topic''} (2/10).

When probed about how they would like prioritize which paragraphs and papers to read, participants mentioned that they wanted to prioritize paragraphs with high coverage, sourced from papers that cover both recent and fundamental work, highly cited and ideally survey papers, and that minimally discuss the paper's own work. Also, in terms of ranking, most (8/10) discussed how the first few paragraphs they read should map out the diversity of subtopics, and (6/10) said similar paragraphs should not be on the same page. 



\subsection{Design Goals} 
Motivated by the findings above, we list our core design goals:

\begin{itemize}
\item[\textbf{[D1]}] Support users in inferring higher-level meaningful organization of topics and dive deeper into subtopics
\item[\textbf{[D2]}] Enable users to fluidly prioritize and explore similarities and differences between related work paragraphs 
\item[\textbf{[D3]}] Help users keep track of paragraphs and references they have explored 
\end{itemize}

\section{The Relatedly System} 
\label{sec: relatedly} 
\begin{figure*}[h!]
\centering
\includegraphics[width=\textwidth]{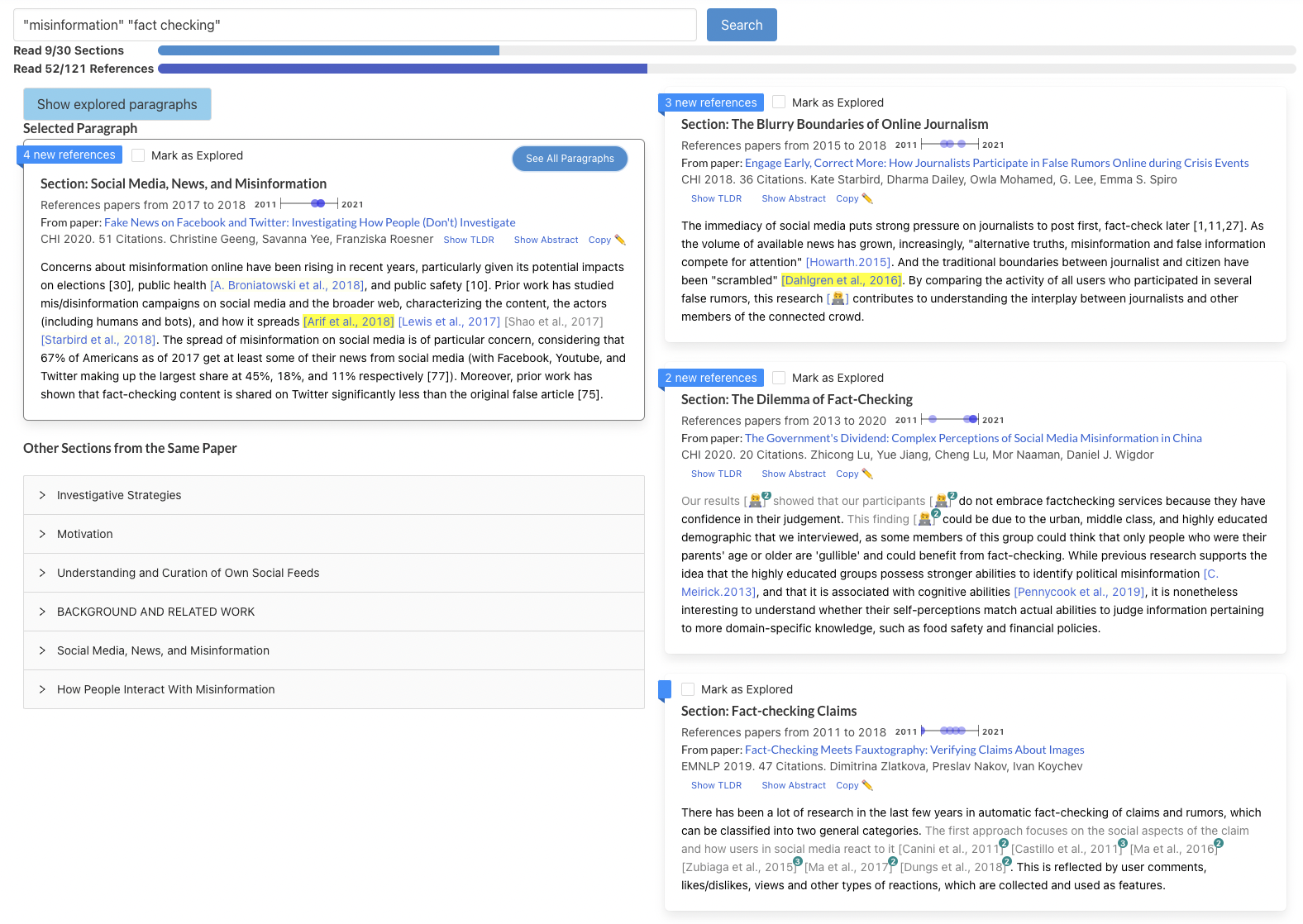}
\caption{To read more on the subtopic discussed in a specific paragraph in the \textbf{Overview View} (Fig.~\ref{fig:Relatedly}), this\textbf{Similar Paragraphs View} allows users to explore other paragraphs of that same subtopic that cited the same or similar references.}
\label{fig:similarParas}
\end{figure*} 

Guided by the insights from our formative study, we developed Relatedly, a novel approach to literature review that helps users achieve a broader and more insightful overview of a research topic. In this section we will describe the system through an example user scenario, a walk-through of the main features of the system, an explanation and evaluation of our automatic section heading generation pipeline, and the implementation details of the system as a whole. 

\subsection{Example User Scenario} 
Consider a junior computer science researcher interested in getting a broad understanding of a research area with which she is unfamiliar---\emph{misinformation and fact-checking}. Not knowing what the important subtopics are, she starts by conducting a literature review using a common scholarly search engine and searches for the phrases: ``misinformation'', ``fact-checking''. However, even though all the papers in the search results look relevant, it is difficult for her to see the higher level themes and how individual papers relate to each other by only looking at the paper titles and search snippets. 

Feeling overwhelmed, she switches to Relatedly with the same query, and the system returns a list of related work paragraphs  relevant to the query in the \textit{Overview View }(Fig.\ref{fig:Relatedly}). Wanting to get an overview, she skims through the section headings and quickly learns different subtopics, such as \emph{Fact Checking Datasets}, \emph{Social Media, News, and Misinformation}, and \emph{Fake News Detection Techniques}. 
The section headings allow her to skim through the Overview View to get a sense of the different high-level research foci. As authors often structure their related works section into relevant subsections based on themes, these titles can help describe the gist of the paragraph's focus. 

As she becomes interested in the subtopic of  \emph{Social Media, News, and Misinformation}, she starts to read the related work paragraph that has it as a section heading. She clicks some of the references to see their metadata, including title, abstract, TLDR \cite{cachola2020tldr}, authors, publication year, conference, and citation count, and she collects some of the ones she wants to read later by clicking ``Copy'' \includegraphics[height=\fontcharht\font`\B]{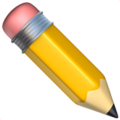}.
Noticing the current paragraph was published four years ago, she clicks on the \textit{``Explore Similar Paragraphs''} button. In the \textit{Similar Paragraphs View} (Fig. \ref{fig:similarParas}), the system brings up other paragraphs from the search result that are also about \emph{Social Media, News, and Misinformation}. 
As she skims the similar paragraphs, she reads about how other author summarized prior work about this particular subtopic across paragraphs (e.g., \textit{The Blurry Boundaries of Online Journalism}, \textit{The Dilemma of Fact Checking}, and \textit{Fact Checking Claims}) extracted from different source papers. She starts to understand connections between multiple referenced papers and concepts discussed in this subtopic. 
She starts to feel like she is getting a more holistic and well-rounded understanding of this subtopic. 
She continues reading other paragraphs including ones that were published more recently to comprehensively explore this subtopic.

She then returns to the Overview View to explore new subtopics. She notices that the paragraphs she is shown have changed as a result of her exploration thus far. Paragraphs have been dynamically re-ranked to prioritize ones with more unexplored and dissimilar references. 
She also notices that some paragraphs have sentences low-lighted that reference papers corresponding to ones she has already explored.
Paragraphs also now sometimes have certain inline citations highlighted that point to unexplored references that are semantically different from those she explored before. She skips over some paragraphs with many sentences low-lighted and focuses on a paragraph with multiple highlights.
Lastly, she checks the progress bar to keep track of what proportion of the entire set of related work paragraphs and references has  she explored thus far.



\subsection{System Features}
We organize the description of the system features according to our three main design goals from the formative study.

\subsubsection{[D1] Infer Topic Overview + Drill-Down to Subtopics} Given a search query, Relatedly presents a list of related work paragraphs relevant to the query in the \textit{Overview View} (\autoref{fig:Relatedly}). Each paragraph is representative of a different subtopic and is ranked to cover a broad range of subtopics.
To drill-down to a subtopic covered by a paragraph (e.g., \textit{Fact Checking Claims}), a user  can click on  the \textit{``Explore Similar Paragraphs''} button on the top right corner of this paragraph card. This brings up the \textit{Similar Paragraphs View} with the similar paragraphs in the right column, and pins the selected paragraph to the left column (\autoref{fig:similarParas}). 
Below are specific features to enable topic overviews and drill-down to subtopics.

\textbf{Diversity ranking of paragraphs}:
To help users see high-level organization in the Overview View, Relatedly first retrieves the most relevant paragraphs based on the standard BM25 \cite{robertson1976relevance} scoring,\footnote{We showed the top 30 paragraphs by default to control for the length of the user study.} and then re-ranks the retrieved paragraphs using the Maximal Marginal Relevance (MMR) technique\cite{carbonell1998use} to balance query-relevance with information-novelty in the top results. 
While the original MMR technique relied on text similarity to measure the information novelty given a document, here we use the number of unexplored references in each paragraph to approximate its information-novelty. The goal is to re-rank the paragraphs such that the top paragraphs jointly contain the most number of unique and unexplored references and present a wide range of diverse subtopics, while accounting for their relevance to the query term and number of references. 
This ranking was determined based on the  participants' responses to how they would like to rank paragraphs in the formative user study. 
The ranking score for paragraph at rank $i$ is as follows:

$MMR_i =  \arg \max_{i} \{ BM25_i [\lambda |{\rm Refs}_i| - (1-\lambda) |((\cup_{j=1}^{i-1}{\rm Refs}_j) \cup {\rm Refs}_{\rm exp})  \cap {\rm Refs}_i|] \}$

In the equation, $BM25_i$ is the relevance score based on the paragraph text and the query term, ${\rm Refs}_i$ is the set of references in the paragraph ranked at position $i$,
${\rm Refs}_{exp}$ is the set of references already explored by the user,
and $\lambda$ is a hyper-parameter for adjusting the penalty for containing references that already appeared in previously ranked paragraphs.\footnote{Based on play-testing during development, we set $\lambda$ to 0.3 to give a moderate advantage to paragraphs containing more references and a high penalty for containing references already covered by higher-ranked paragraphs to diversify topic coverage of the top results.} 


\textbf{Descriptive paragraph headings}:
Each paragraph in the Overview View comes with a descriptive title to describe the gist of the paragraph's focus and serve as a subtopic for the user. 
As authors often structure their related works section into relevant subsections based on themes, for most paragraphs, these are extracted from the section headings of the related work sections. 
For paragraphs for which authors had only written generic or short section headings that contained less information (e.g. ``Related Work'' or ``Fact-Checking''), or for paragraphs with no section headings, Relatedly generates descriptive titles using a BART-based \cite{lewis-etal-2020-bart} model (described in more detail in §\ref{sec:autogeneratedTitles}).

\textbf{Similar paragraphs given a paragraph}:
In the Similar Paragraphs View, a list of similar paragraphs are shown for a selected paragraph.
We determine paragraph similarity by whether they reference either the same papers as the selected paragraph (primary sort-order), or are semantically similar papers (secondary sort-order, using a threshold on the Euclidean distances between their SPECTER paper embeddings \cite{cohan2020specter}).

\textbf{Reading the paper behind a given paragraph}:
In the Similar Paragraphs View, underneath the selected paragraph, the user can access all the other sections from the same paper, including other portions of the related work section, in order to gain more context behind the paragraph.


\subsubsection{[D2] Prioritize and explore similarities and differences across related work paragraphs}
In the formative study, participants expressed that it was ``hard to prioritize in which order to read the paragraphs''. 
Thus, Relatedly presents a number of features to support prioritizing and reading diverse unexplored information.

\textbf{Unexplored references count badge}:
Both our formative interviews and prior work pointed to wanting to prioritize paragraphs that had the highest unread information first. To aid with this, all paragraphs have an unexplored references count badge (like \includegraphics[height=2\fontcharht\font`\B]{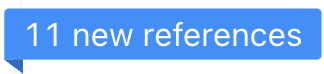}) that conveys the number of unique unexplored number of papers discussed in this paragraph. This number dynamically updates as the user interacts with more references across the paragraphs. The ranking algorithm prioritizes and ranks paragraphs with more unread references higher in the Overview View. 

\textbf{Highlighting of dissimilar unexplored references}:
To further facilitate prioritizing unexplored novel information and address the need to \textit{``identify similarities and differences between papers''}, 
Relatedly highlights dissimilar unexplored references (like  \includegraphics[height=1.5\fontcharht\font`\B]{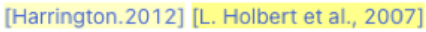}). As the user clicks and reads references and paragraphs, some references get highlighted yellow indicating that these papers are semantically different to other papers interacted with so far (calculated using a threshold on
the Euclidean distances between their SPECTER paper embeddings \cite{cohan2020specter}). These references are highlighted on a yellow gradient, where the brighter the yellow, the highlighted paper is more different than the most similar papers interacted with so far. 

\textbf{Reference timeline visualization}:
Another heuristic that users wanted to use was to prioritize paragraphs that covered both recent and fundamental prior research on the topic. To help triage this, all paragraphs have a  reference timeline visualization that visualizing the time-range of papers referenced in this paragraph (like \includegraphics[height=1.5\fontcharht\font`\B]{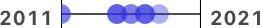}). Here, each semi-transparent blue dot is a referenced paper publication year. As the min and max of the timeline signify the earliest and latest papers referenced across all paragraphs, users can use this to prioritize reading paragraphs that reference recent papers or more fundamental older papers. This feature was based on the formative study participants' saying they wanted to triage reading priority based on the recency of papers referenced in the paragraphs.

\textbf{Citation frequency badges}:
Participants in the formative interviews mentioned wanting to prioritize parts of a paragraph, particularly wanting to know which paper to prioritize when there are multiple papers cited for a claim or in a paragraph. Relatedly offers Citation frequency badges that aim to indicate how central or important a referenced paper is to a topic. These green tags with a number refer to the number of times this paper has been referenced in these result paragraphs (like  \includegraphics[height=1.8\fontcharht\font`\B]{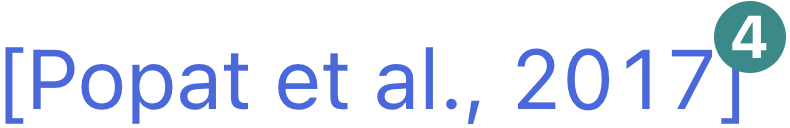}). If more than one of the paragraphs returned in the Overview View referenced it, it means that multiple authors discuss this paper, therefore it might be central or important to this topic. 

\textbf{Self-reference icons}: Participants in the formative interviews mentioned wanting to de-prioritize parts of a paragraph where the authors were situating their own work in the background. To aid this Relatedly identifies which parts of the paragraph refer to the paper's work and signal this to the reader with an \includegraphics[height=2\fontcharht\font`\B]{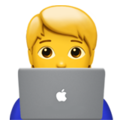} icon.


\subsubsection{[D3] Keeping Track Exploration Progress over Papers and Paragraphs}
In addition to wanting to prioritize dissimilar unexplored information, users mentioned that it was challenging to track which papers or paragraphs have been read versus not. Relatedly provides a number of features to give users a sense of their progress in covering content while minimizing redundancy.

\textbf{Low-lighting previously encountered information}
Relatedly low-lights previously encountered information by graying out the entire sentence in a paragraph. If a user has clicked on a referenced paper in a paragraph and it is referenced in another unseen paragraph, the reference and the corresponding text will be low-lighted there too to indicate that they have previously encountered this information. 

\textbf{Mark paragraphs as explored}:
Similarly, at the paragraph level, once a paragraph is \textit{``marked as explored''}, it is removed from the Overview View. To access these explored paragraphs, a user can click on the \textit{``Show explored paragraphs''} button at the top left of the page. Also, every time a paragraph is marked as explored, all of its references are considered as ``encountered'' and there is a dynamic re-ranking of the paragraphs list based on the number of unexplored papers in them, how dissimilar the paragraph is to what has been read, and the relevance to the topic queried (formula 1, described in a previous section). 

\textbf{Progress bars}:
As paragraphs get marked as explored, the paragraph progress bar at the top of the page is updated (like \newline 
\includegraphics[height=\fontcharht\font`\B]{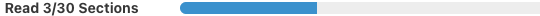}...).
As the user clicks on references, the paper progress bar at the top of the page gets updated as well and conveys that the user has read \textit{n} out of the total number of unique references across the paragraphs returned (like \includegraphics[height=1\fontcharht\font`\B]{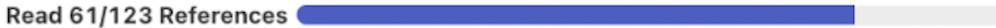}). The  unexplored reference count badges across paragraphs also get updated. This remains persistent across queries, so as a user issues new queries and if they have read any of the papers or paragraphs before, these would would be tracked in the progress bar too. This feature is designed to help address the user challenge that it is difficult to keep track of information read over papers and subtopics explored.

\subsection{Automatic Section Heading Generation}
\label{sec:autogeneratedTitles}
Relatedly leverages section headings of related work sections to provide users a quick overview over different research threads. 
One challenge here is that not all authors create descriptive subsection headings. In these cases, showing the generic higher-level section heading (e.g., ``Related Work'' or ``Fake News'') for the query of \emph{fake news} provides little value to the users. To address this, we developed an automated heading generation model, trained on heuristically identified descriptive section headings, and applied it to paragraphs that did not have descriptive headings written by the authors of the source paper.

\subsubsection{Method}
We experimented with two popular transformer-based sequence-to-sequence models for heading generation: (i) BART \cite{lewis-etal-2020-bart}, and (ii) T5 \cite{raffel2020exploring}.
These pre-trained models have become de-facto starting points to develop various text generation models due to their strong performance and ability to adapt to different tasks.
To further train these models for scientific heading generation, we use a set of heuristics to gather paragraphs that contain descriptive titles from our dataset. These heuristics include filtering out all titles that are single acronyms, shorter than three words, or contain generic terms.\footnote{literature review, background, limitations, future work, conclusion, discussion, related work, results.} This strict filtering favors precision over recall in order to reduce the amount of noise in the filtered dataset for training and testing. Our final dataset consists of 23,957 paragraphs and their titles, which we further randomly divide into 80\% training, 10\% validation and 10\% test splits.
We train the large variants of both BART and T5 on this training split for 10 epochs and use loss on the validation split to select the best-performing model checkpoint. Table~\ref{tab:rouge} presents an evaluation of both models on our held-out test split using the ROUGE metric \cite{lin-hovy-2002-manual}, which measures title quality via n-gram overlap between generated and human-written titles. Based on these scores, the two models seems to generate similar-quality titles, and we sampled and examined a small subset of the generated headings to compare the two models and found that the BART model tend to produce headings that were more detailed and descriptive. Therefore, we then used the BART model to generate headings for paragraphs outside of the filtered set that did not have descriptive author-generated headings (39,186 in total or around 62\% of the entire dataset) for a more rigorous human evaluation detailed in the next subsection.

\begin{figure*}
\centering
\includegraphics[width=0.65\textwidth]{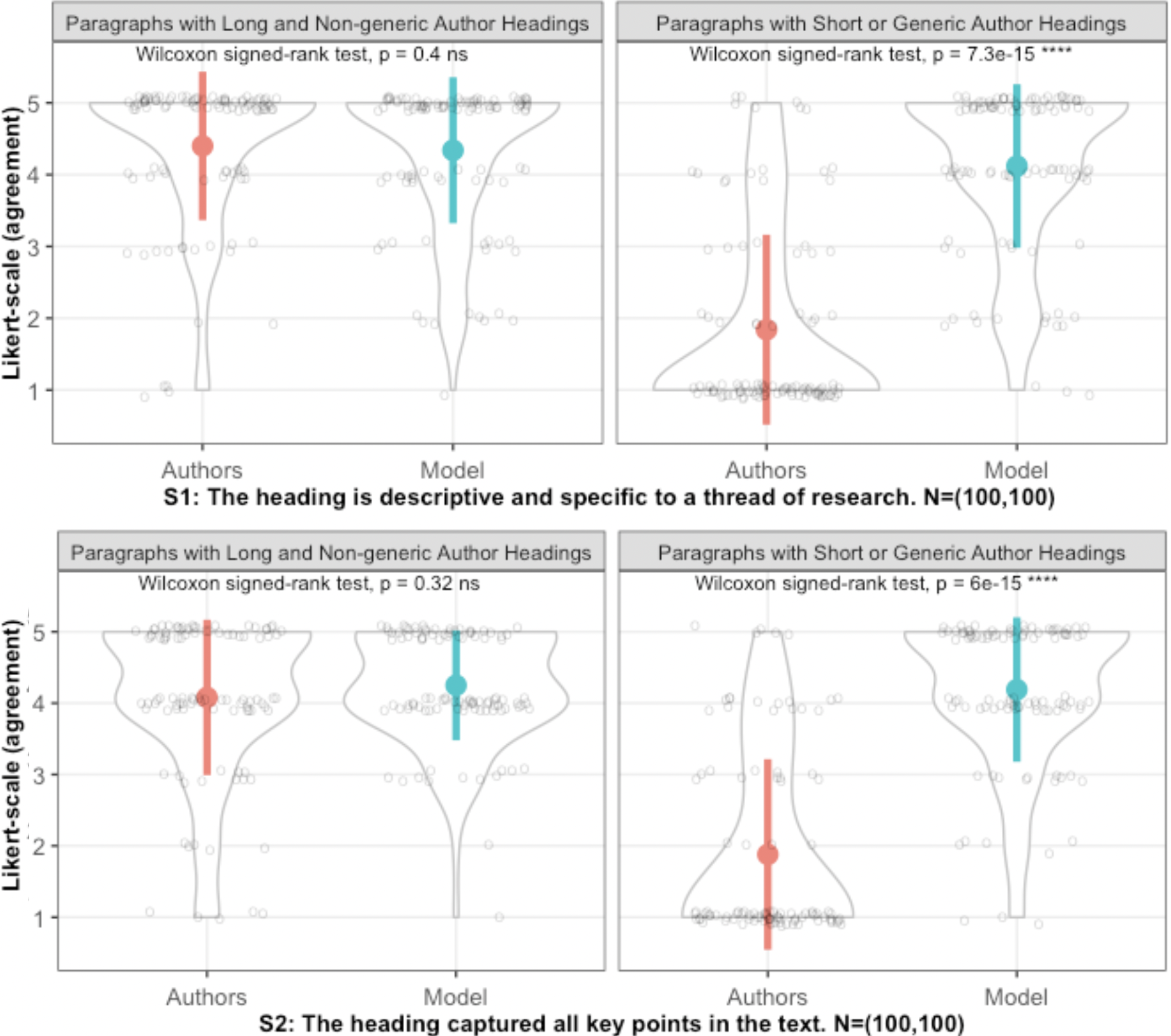}
\caption{Human-evaluation comparing model-generated and author-written section headings. Results suggest that model-generated headings were of comparable quality when the authors had written long and non-generic headings and were of significantly higher quality when the authors did not.}
\label{fig:headingGen}
\end{figure*} 

\begin{table}[]
    \centering
    \small
    \begin{tabular}{lccc}
    \toprule \textbf{Model} & \textbf{ROUGE-1} & \textbf{ROUGE-2} & \textbf{ROUGE-L} \\ \midrule
    \textbf{BART} & 30.7 & 14.2 & 28.8 \\
    \textbf{T5} & 31.0 & 14.3  & 29.0 \\ \bottomrule
    \end{tabular}
    \caption{ROUGE scores for both models on the test split of descriptive section headings. ROUGE-1, -2, and -L measure unigram, bigram, and longest subsequence overlap between generated and author-written titles, respectively.}
    \label{tab:rouge}
\end{table}

\subsubsection{Human Evaluation}
We conducted a human evaluation of the BART-based heading generation model. This is important because automatic metrics, such as ROUGE, do not always correlate well with human perception. We manually rated four sets of headings: both author- and model-generated headings for both paragraphs with descriptive headings but not included in the training (the test split) and paragraphs without descriptive headings. Two statements were rated for 5-point agreement: \emph{The heading is descriptive and specific to a thread of research} (S1) and \emph{The heading captured all key points in the paragraph} (S2). S1 aimed to measure the quality and specificity of the headings, and S2 aimed to measure how well they represent the paragraph text.

To ensure rating quality, two of the authors went through two rounds of redundant rating of 40 randomly sampled headings per round (10 from each set). After two rounds of comparison and discussion to calibrate rating standards, inter-annotator agreements based on Krippendorff's alpha reached 0.90 and 0.74 for S1 and S2, respectively.\footnote{Agreements from round 1 were 0.66 and 0.49 for S1 and S2, respectively.} The authors then proceeded to rate 400 headings (100 from each set, paired) without redundancy. During the rating process the authors were blind to the condition each heading was sampled from to avoid bias.

As shown in Fig.~\ref{fig:headingGen}, our model was able to generate headings of comparable quality to long and descriptive titles written by the authors (S1: p=0.40; S2: p=0.32; $n=[100,100]$, Wilcoxon signed-rank tests), and significantly higher quality headings when descriptive headings were not available from the authors (S1: p<0.001{$^{***}$}; S2: p<0.001{$^{***}$}; $n=[100,100]$, Wilcoxon signed-rank tests). In addition to these ratings, we also quickly screened all model-generated titles (200 in total) for repetition and hallucinations, i.e. mentions of concepts not present in the paragraph. We do this as a sanity check since generation models are often prone to these issues, especially hallucinations \cite{ji2022survey}. Based on our screening, we found that our model rarely suffers from these issues - only 5/200 (2.5\%) titles have repetition and 9/200 titles (4.5\%) have hallucination. This may partly be due to the fact that generated headlines are fairly short, which offers less scope for repetition and hallucination to creep in.  
Given these promising results, in the system, we showed model generated titles whenever a paragraph did not have a descriptive author-written heading (Fig.~1). Some examples of model-generated headings side-by-side with their corresponding author-written headings are presented in the Appendix (Table~\ref{tab:tab_long_heading} and ~\ref{tab:tab_short_heading}).

\subsection{Implementation Details}
Relatedly is built as a standard web application. The front-end was written in approximately 3,500 lines of TypeScript using the ReactJS framework. The back-end is implemented in approximately 1,500 lines of Python and SQL code. We used Flask for HTTP server framework and PostgreSQL database for both dataset access and behavior logging for the user studies. We used the Whoosh\footnote{\url{https://github.com/mchaput/whoosh}} Python library, which implements the standard BM25 document retrieval algorithm \cite{robertson1976relevance}, to support full-text search of the paragraphs. For the evaluation study, all interactions with the system (such as new queries, papers and paragraphs read, etc.) are logged to a database in a JSON format (refer to supplementary materials). All communications between the server, database and users’ browser are encrypted and anonymized by creating anonymous session and user IDs.

\subsubsection{Dataset}
To test the Relatedly approach, we gathered a dataset of full papers from five HCI and NLP conferences (ACL, EMNLP, UIST, CSCW, CHI) published between 2016-2021 from S2ORC, a large open-source corpus of 81.1M English-language academic papers spanning many academic disciplines \cite{lo-etal-2020-s2orc}. These topics were selected out of convenience so that the authors could evaluate the usefulness of the system during development, and also for recruiting participants who are likely more engaged with this topic during our user studies.
To find paragraphs that summarize multiple prior studies, for each paper, we extracted all paragraphs that contained three or more references along with their section titles. Since a related work section would typically reference its source paper, which can seem out of context when read independently, we used a simple word list to resolve self-referencing phrases (e.g., \emph{in this paper}, \emph{our approach}, \emph{our system}, ..., etc.) to the source paper.
In the end, this dataset contained 63,144 paragraphs extracted from 11,382 papers. Approximately 49,975 paragraphs were from related work sections and the remaining paragraphs were mostly extracted from introduction and discussion sections. The inline references were resolved by S2ORC \cite{lo-etal-2020-s2orc} to their metadata including authors, citation count, abstract, and TLDRs \cite{cachola-etal-2020-tldr}. This also allowed \textit{Relatedly} to reformat the reference text into APA format (i.e., the first author's last name and the publication year) in the system so that the same references have the same surface form across paragraphs. 

\section{User Evaluation Study Design}
\label{sec:method}
\begin{figure*}[!h]
\centering
\includegraphics[width=\textwidth]{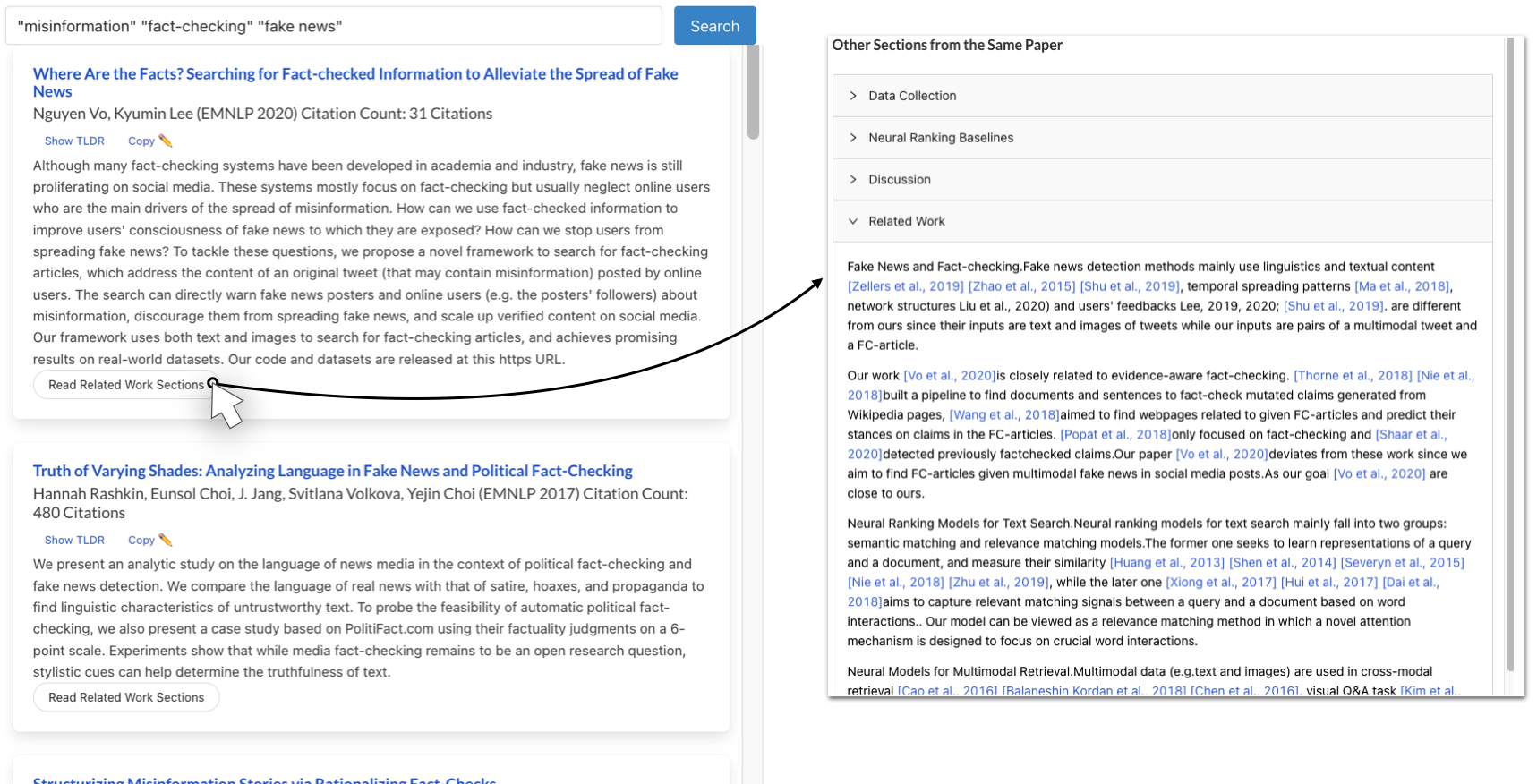}
\caption{The Baseline condition that emulates common scholarly search engines (left). In addition, the ``Read Related Work Sections'' buttons allow users to read the paper's sections with paragraphs that contained three or more references with lowered-interaction costs (right).}
\label{fig:Baseline}
\end{figure*} 


The design of \textit{Relatedly} changes the common literature review process of exploring individual papers (e.g., from a search engine), to exploring paragraphs describing multiple papers on a topic. To investigate its effects, we conducted a within-subjects experiment with 15 participants conducting literature reviews comparing \textit{Relatedly} to a standard paper search engine as baseline. During the study, participants used the assigned system to explore the literature
with the goal of creating an outline and notes for writing a survey paper on assigned topics. This allowed us to capture what participants had learned during the tasks. After the study, we analyzed the behavior logs to understand how they utilized each system, and rate the quality of their outlines to see which process allowed participants to gain a better overview of the literature.

\subsection{Experimental Setup} 
We compared \textit{Relatedly} to a Baseline system simulating standard scholarly search engines as an within-subject condition. During the study, each participant used both systems to conduct literature reviews on two different topics. To control for individual differences and learning behavior, we counterbalanced topics and conditions to reduce order effects. 

\subsubsection{The Baseline Conditions} 
The baseline condition was a standard BM25 search engine that returned a list of papers from our dataset that mentioned the query term in their titles or abstracts (Fig. ~\ref{fig:Baseline}). For each paper in the search results users can access its metadata including the title, authors, venue, publication year, abstract, and a TLDR summary\cite{cachola2020tldr}.\footnote{These are 1-2 sentence summaries also available on popular scholarly search engines.} 
To lower the interaction costs of using the Baseline condition, users can click on a ``Read Related Work Sections'' button to access the section headings and paragraphs that contained three or more references (Fig~\ref{fig:Baseline}). This ensures that 1) participants have access to the same data in both conditions, and that 2) the interaction cost of accessing them is low in the Baseline. 
Similar to the \textit{Relatedly} condition, participants could also click a ``copy'' button to copy a paper title and paste into their outlines. 


\subsubsection{Tasks} 
To contextualize their exploration and sensemaking, participants were given a simulated work task scenario \cite{borlund2003iir} to conduct initial research to get a broad overview of the topic towards the ultimate goal of writing a survey article: 
\begin{quote}
Imagine that you are surveying and summarizing scientific work in HCI and NLP on the topic of: 

[\textit{One of two task topics:  Misinformation, Fake News and Fact Checking OR Crowdsourcing}]

Today, please do an initial research to get a broad overview of the topic. Your goal should be to get a broad overview of this topic and identify as many terms, concepts and perspectives related to the topic as you can find by searching and gathering information on this search engine. During the task, write an outline in the notes document provided to you such that it would help you resume work on this task in the future. This may include planning out all the sections of your paper, recording important papers and research you find, etc.
\end{quote}

As part of the within-subjects study design for evaluating user behavior across the two conditions, each participant worked on the above task twice (i.e. once for each condition). To prevent carryover effects in learning, each participant completed the task on the two topics listed below. To avoid order effects, the systems were counterbalanced such that they saw a different topic with each condition. 

\begin{itemize}
    \item \textbf{Misinformation, Fake News and Fact Checking:} The internet makes it easy for billions of people to access information with a few simple keystrokes. However, it also makes it easy to spread false information, which can have disastrous effects on both individuals and society as a whole. Research in HCI and NLP has focused on detection methods, their use and impacts. Research the impacts of fake news and the methods being developed to combat it. 
    
    \item \textbf{Crowdsourcing:} Crowdsourcing involves a large group of dispersed participants contributing to a task. As we move towards a new future of work with digital platforms for crowdsourcing, research in HCI and NLP has focused on the different methods of crowdsourcing and the applications across different domains. Research the methods and applications of crowdsourcing.  
\end{itemize}

The chosen task topics are relatively large, complex, multi-faceted information spaces where the average person has relatively limited knowledge coming into the task. They are also fairly interdisciplinary tasks so that even if we do get people with domain expertise, there’s more they can learn in this area. Also, these topics are well-represented in our dataset -- which is papers for HCI and NLP conferences.

\subsubsection{Participants} 
15 participants were recruited from research labs across three universities (8 identified as female and the rest as male; age: 19-32. M=25.88, SD=3.72).
All studies were conducted remotely over video calls. Compensation was \$45 USD for the 90 minute study. The participants were mostly research scientists, post-docs, and graduate students engaging in research activities.

\subsection{Study Procedure} 

Before the study appointment, participants were sent the informed consent form and asked to fill out demographic information. During their study appointment, each participants went through two literature review tasks where the order and the combination of tasks and system assignments were counterbalanced. Each of the two tasks lasted 20 minutes. During the 20 minutes, participants freely interact with the system and create their learning outline on a Google Doc while thinking outloud about their experiences \cite{lewis1982using, world_leaders_in_research_based_user_experience}. 
Before starting each task, participants watched a short tutorial on each system and were given 5 minutes to explore the system using the test topic of ``sensemaking''.

\subsection{Measures} 

\subsubsection{\textbf{Quality of Learning Outlines}}
Our primary measure focused on how well \textit{Relatedly} supports literature reviews compared to the baseline by analyzing the learning outline participants wrote in the lab study. For this, we used topic experts to examined and rate each of the Google Doc outlines while being blind to which condition they came from. We defined an expert as someone who has obtained a doctoral degree focusing on the topic, and had multiple years and publications in the field. Two of the authors matched these criteria for the tasks used in the study, blind to condition. The experts counted the number of research themes participants added to their outlines (a proxy for \emph{comprehensiveness}), and rated the outlines for the following aspects on a five-point scale (higher is better): 
\begin{itemize} 
\item \textit{Coherence}: The category structures make sense and the papers and subcategories in the them fit.
\item \textit{Insightfulness}: The categories were insightful and captured important research threads in the space.
\item \textit{Level of Detail}: The categories contain rich details of relevant subtopics and papers.
\end{itemize} 
These criteria were inspired by literature on human evaluation of clustering (e.g. \cite{zhang2018evaluation}) and NLP evaluation criteria for automatically generated outlines (e.g. \cite{gehrmann-etal-2019-improving}) The two experts went through two rounds of rating and discussion to calibrate the number of themes and their final scores.
The sum of these scores is then used to calculate overall quality of the outline. 

\subsubsection{\textbf{Behavior Log Analysis}} Using the behavior logs from both conditions, we measured how frequently each participant interacted with both the systems at both the paragraphs and papers level. For example, references clicked on, paper titles copied, or paragraphs explored. In addition, we also examined how frequently participants interacted with features only available in the \textit{Relatedly} condition, such as reference low- and high-lights, citation frequency badges, and using the Overview View and Similar Paragraphs View.


\subsubsection{\textbf{Qualitative Insights and Perceived Values}} 
In order to gain deeper understanding of the challenges and benefits of using the two systems, 
we transcribed participants' think-aloud recordings during the tasks. The first author then went through the transcripts in two passes using an open coding approach \cite{charmaz2014constructing}. Through discussions with the rest of the research team, we identified common themes in participants' experiences. Additionally, we also conducted a post-task survey where we asked participants to rate a set of statements around system values (Table 3) using a 5-point Likert-scale for agreement.

\section{Findings}
\label{sec:results}
\begin{figure*}[ht!]
\centering
\includegraphics[width=\textwidth]{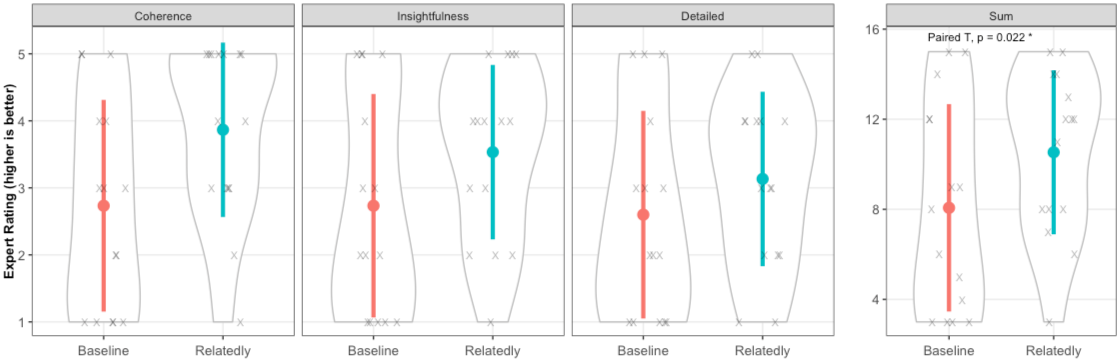}
\caption{Participants generated learning outline summaries after 20 minutes of literature review each with the two systems. The summaries were rated by experts on 3 criteria using
5-point Likert-scales for agreement (5 indicated strong agreement). The experts were blind to which condition each outline came from during rating. A MANOVA and a Wald-type test were used to correct for multiple comparisons and found a statistically significant difference ($manova F=7.78, p=0.02^*, Wald \chi^2=13.18, p=0.004^{**}$) between the conditions on the combined dependent variables
of Coherent, Detailed, and Insightful. }
\label{fig:Synthesis}
\end{figure*} 
In this section, we combine results from our three measurements described in the previous section to give a holistic view of the costs and benefits of using \textit{Relatedly} when compared to the Baseline condition, which simulates standard scholarly search engines.


\subsection{Higher Quality Synthesized Outlines}

Based on the sum of 5-point expert ratings on the three aspects, participants wrote significantly \textit{better quality outlines} when using Relatedly compared to the Baseline ($M=10.53, SD=3.64$ vs $M=8.07, SD=4.61; t=2.58, p=0.02$ out of 15, Fig. \ref{fig:Synthesis}). To correct for multiple comparisons, we ran a Wald-type test and a MANOVA with repeated measures and found a significant difference between the two systems on the combined measures of \emph{Coherence}, \emph{Insightfulness}, and \emph{Detailed} ($manova F=7.78, p=0.02^*, Wald\chi^2=13.18, p=0.004^{**}$; Fig~\ref{fig:Synthesis}).\footnote{R package: \emph{MANOVA.RM: Resampling-Based Analysis of Multivariate Data and Repeated Measures Designs
}} 
Breaking this quality score down, participants wrote significantly more coherent ($M=3.87, SD=1.30,$ out of 5) and insightful ($M=3.53, SD=1.30,$ out of 5) outlines when using Relatedly compared to when using Baseline (Coherence: $M=2.73, SD=1.58$ out of 5; Insightfulness: $M=2.73, SD=1.67$ out of 5). Participants also wrote more detailed outlines when using Relatedly ($M=3.13, SD=1.30$ out of 5) compared to when they used Baseline ($M=2.60, SD=1.55$ out of 5). 



Qualitative analysis of the think-aloud recordings revealed participants' exploration strategies when using \textit{Relatedly}. Most commonly, 13 out of the 15 participants talked about using a ``breadth-first approach for exploring different paragraphs and topics,'' and 8 specifically mentioned using the Unexplored Reference Count Badge (\includegraphics[height=2\fontcharht\font`\B]{Images/ribbon.png}) at the paragraph level to prioritize. For example, P09 used the Overview View to quickly capture diverse topics and relevant papers in their outline, taking advantage of how \textit{Relatedly} ranked the paragraphs dynamically to maximize marginal novelty \cite{carbonell1998use}:

\begin{quote}
\textit{``I spent most of my time in the all paragraphs view looking at the various summaries. -- I like that they are in [the] order of most unread references to fewest, so it felt like going in-order made sense. As I found new topics, I jotted them down -- sometimes as a short summary, sometimes I included entire quotes of the overall message, and then copied over the related paper to go back to reference if I needed.''} (P09).
\end {quote}

\noindent All participants switched between the Overview View and Similar Paragraph View for broad overview and drill-down into different subtopics. Furthermore, when using Relatedly, participants explored significantly more paragraphs in the Overview View ($M=5.80, SD=8.00$) than in the Similar Paragraphs View ($M=1.80, SD=1=.13, t=2.55, p=0.02^{*}$). Two participants specifically mentioned switching between the two views to control both the depth and breath of their explorations:

\begin{quote}
\textit{``I'd describe my searching as like a limited-depth depth first search strategy. I start at a high level idea, try to find everything related to that topic up to some effort level. Similar Paragraphs [View] was a great feature for this! Then, I switch to another sub topic [in the Overview View] and repeat.''}  (P17). 
\end{quote}

\begin{figure}
\centering
\includegraphics[width=\columnwidth]{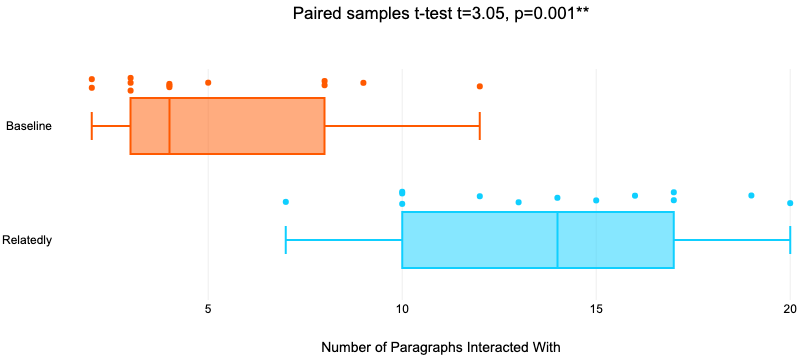}
\vspace{-5mm}
\caption{Participants interacted with significantly more paragraphs when using Relatedly vs the Baseline system}
\label{fig:ParasResult}
\end{figure} 





These system log analysis and qualitative insights suggested that participants had more control over their exploration when using \textit{Relatedly}, fluidly switching between exploring many diverse subtopics and drill-down to specific subtopics, and are potential explanations for how they collected more subtopics and generated higher quality overview learning outlines.

\subsection{Paper- vs Topic-Centric Exploration}

While participants in both conditions had access to papers' related work paragraphs, we found that when using \textit{Relatedly},
participants interacted with more than twice as many paragraphs compared to when they were using the Baseline (Relatedly: $M=16.85, SD=7.66$ vs $M=7.21, SD=5.70, t=4.40, p=0.001^{**}$). 
Participants described their exploration strategies were centered around individual papers instead of paragraphs when using the Baseline. Most commonly, nine (/15) mentioned relying mostly on the abstracts and TLDR summaries to decide which papers to read: \textit{``Go paperwise, skim through abstract - read more if it's interesting or relevant"} (P04). Participants also mentioned using other signals such as the citation counts (3/15), spotting unfamiliar terms in the abstracts to find categories to add to their outlines (3/15), and reading related work sections (4/15). 

In addition to differences in exploration strategies between the two conditions, we also found participants actively used the paragraph reading support features when using \textit{Relatedly} based on both behavioral and qualitative data.
On average, 38.73\% of citations clicked were low-lighted previously read references, 37.84\% of citations clicked were highlighted and 10.35\% of citations clicked had citations frequency badges. 
Based on qualitative data, participants utilized the reading support cues to both prioritize and de-prioritize their reading activities. 12/15 participants talked about relying on the yellow highlighted references to prioritize their reading: 
\begin{quote}
    ``\emph{oh I see yellow, which means that there is something different, and the brighter the yellow, it looks more attractive to me. I want to see papers that disagree to the papers I’ve read so far}'' (P04).
\end{quote}
\noindent In addition, 11/15 participants paid attention to the citation frequency tags to find important papers on the topic: 
\begin{quote}
``\emph{I will look for the citation with high numbers, because those tend to be popular and more classic or fundamental.}'' (P03). 
\end{quote}
For deprioritization, 12/15 participants talked about their use of low-lighted references:
\begin{quote}
``\emph{when I start to read a paragraph and I see grayed out text and references, it is very helpful to see that I have read about this paper before in another paragraph. I'm going to click on this to verify which paper this is and then read how this paragraph is discussing it. Oh it seems to be a slightly different take on this paper's contributions so I'll copy this paper. It seems like there's some discussion around it.}'' (P09). 
\end{quote}

Finally, participants mentioned using the descriptive section headings to explore topics when using \textit{Relatedly}. Ten participants explicitly mentioned how the section titles indicate topics that are useful for organizing knowledge: 

\begin{quote}
\textit{``These [titles] are roughly the broad categories for which I would look for, and I’m first going through the titles and adding the unique ones to my notes … I will now start looking into specific things similar to this subtopic title''} (P14). 
\end{quote}

\subsection{Participants Preferred Relatedly}
\begin{figure*}
\centering
\includegraphics[width=\textwidth]{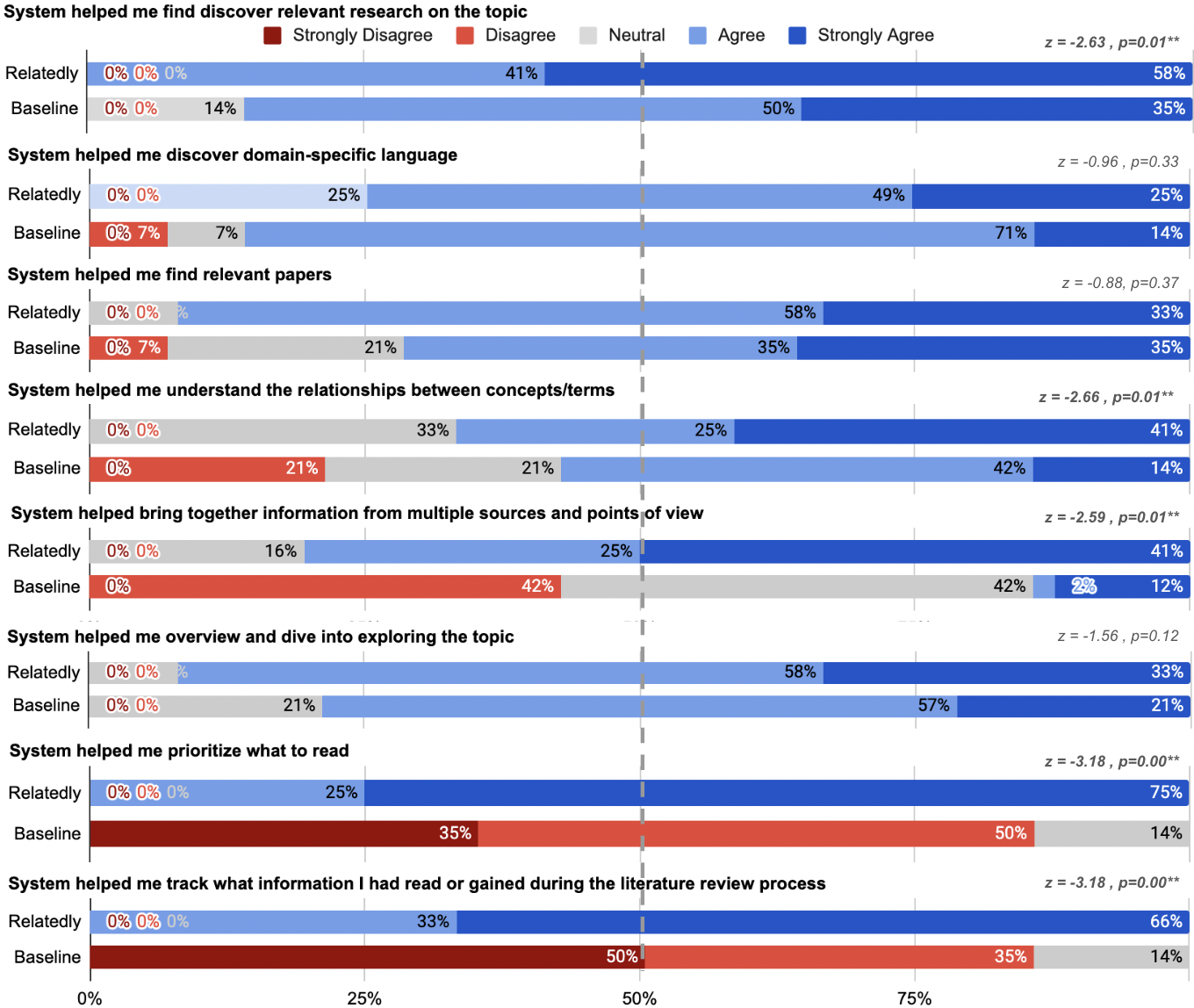}
\caption{Participants' level of agreement to how well Relatedly and Baseline supported their literature review process. For each statement, we report the percentage of likert responses and results from paired wilcoxon signed rank tests, with z and p values.}
\label{fig:Prefs}
\end{figure*} 

After using both systems, participants were asked about their preferences and 13 / 15 participants said they preferred \textit{Relatedly's} approach of reading related work paragraphs on the topic to the Baseline's approach of reading papers to review literature. While the two participants who preferred the Baseline mostly described it as a familiar interface, participants who preferred \textit{Relatedly} reported much richer explanations:
(1) provides a good structure and organization to an otherwise unstructured complex exploratory task: \textit{``I can have a relatively clear path to explore. I can use what other researchers have summarized so I don't need to start from zero.''} (P05); (2) helps understand connections between multiple papers: \textit{``For literature review, it is important to see connections between cite works... I find the Related work sections helpful for this, not so much for the paper list.''} (P14); (3) gives the right amount of relevant context around papers: \textit{``Gave a lot more context and resulted in fewer papers being read -- had to open fewer pdfs.''} (P13); and (4) helps track progress and prioritize what order to read things in:  \textit{``can help me to keep track of my pace of learning about the topic."} (P03). 


In a post survey about participants' opinions about the two systems, participants thought the Relatedly system helped them significantly more than the Baseline for "find relevant research on the topic", "understand relationships between terms/concepts", "bring together information from multiple sources and points of view", "prioritize what to read" and "keep track of information gained or read during the literature review process" (Wilcoxon signed-rank tests for all statements reported in Fig. \ref{fig:Prefs}). 

Lastly, to measure the perceived workload of using the two systems, we used the NASA TLX questionnaire, and found no significant differences (Appendix Table \ref{tab:nasatlx}). This result suggests that participants did not perceive higher workload when using \emph{Relatedly} even though it consisted of significantly more features, and that they were able to utilize these features to synthesize learning outlines of significantly higher quality.

\subsection{Volunteered Use in the Wild}
\begin{table*}[]
\small
\begin{tabular}{lllllll}
{\color[HTML]{000000} \textbf{PID}} & {\color[HTML]{000000} \textbf{Job Title}} & {\color[HTML]{000000} \textbf{\begin{tabular}[c]{@{}l@{}}Reason for using \\ Relatedly during their\\  research workflow\end{tabular}}} & {\color[HTML]{000000} \textbf{\begin{tabular}[c]{@{}l@{}}Topic of\\ Interest\end{tabular}}} & {\color[HTML]{000000} \textbf{\begin{tabular}[c]{@{}l@{}}Hours\\ using \\ Relatedly/ \\ Total Hours\\ Working\end{tabular}}} & {\color[HTML]{000000} \textbf{\begin{tabular}[c]{@{}l@{}}\# of\\ queries\end{tabular}}} & {\color[HTML]{000000} \textbf{\begin{tabular}[c]{@{}l@{}}\# of \\ papers\\ curated\end{tabular}}} \\
{\color[HTML]{000000} P07} & {\color[HTML]{000000} \begin{tabular}[c]{@{}l@{}}Research \\ Assistant\end{tabular}} & {\color[HTML]{000000} \begin{tabular}[c]{@{}l@{}}Finding and curating \\ relevant papers into\\ an annotated reading list\end{tabular}} & {\color[HTML]{000000} \begin{tabular}[c]{@{}l@{}}Smartphone accessibility\\ techniques for people with\\ motor impairments\end{tabular}} & {\color[HTML]{000000} 4/4} & {\color[HTML]{000000} 10} & {\color[HTML]{000000} 30} \\
{\color[HTML]{000000} P09} & {\color[HTML]{000000} \begin{tabular}[c]{@{}l@{}}Post-\\ Doctoral\\ Researcher\end{tabular}} & {\color[HTML]{000000} \begin{tabular}[c]{@{}l@{}}Researching related \\ work to identify gaps\end{tabular}} & {\color[HTML]{000000} \begin{tabular}[c]{@{}l@{}}Cognitive and design \\ theories on feedback\end{tabular}} & {\color[HTML]{000000} 25/36} & {\color[HTML]{000000} 14} & {\color[HTML]{000000} 102} \\
{\color[HTML]{000000} P11} & {\color[HTML]{000000} PhD student} & {\color[HTML]{000000} \begin{tabular}[c]{@{}l@{}}Starting a new project in\\ an unknown research topic\end{tabular}} & {\color[HTML]{000000} \begin{tabular}[c]{@{}l@{}}Creativity support tools,\\ for 3D prototyping\end{tabular}} & {\color[HTML]{000000} 10/10} & {\color[HTML]{000000} 4} & {\color[HTML]{000000} 98} \\
{\color[HTML]{000000} P13} & {\color[HTML]{000000} \begin{tabular}[c]{@{}l@{}}Machine \\ Learning \\ Researcher\end{tabular}} & {\color[HTML]{000000} \begin{tabular}[c]{@{}l@{}}Checking for unknown \\ papers when writing a \\ paper on a known topic\end{tabular}} & {\color[HTML]{000000} \begin{tabular}[c]{@{}l@{}}Multi-document\\ summarization techniques\end{tabular}} & {\color[HTML]{000000} 5/5} & {\color[HTML]{000000} 3} & {\color[HTML]{000000} 51} \\
{\color[HTML]{000000} P15} & {\color[HTML]{000000} \begin{tabular}[c]{@{}l@{}}Research \\ Scientist\end{tabular}} & {\color[HTML]{000000} \begin{tabular}[c]{@{}l@{}}Checking for unknown \\ papers when writing a \\ paper on a known topic\end{tabular}} & {\color[HTML]{000000} \begin{tabular}[c]{@{}l@{}}Use of AR/VR techniques\\ in health and education\end{tabular}} & {\color[HTML]{000000} 1.5/4} & {\color[HTML]{000000} 11} & {\color[HTML]{000000} 23}
\end{tabular}
\caption{Background and usage of participants  who volunteered long-term use. Participants self-reported their job title, reason for using Relatedly in their research workflow, hours of work, hours using Relatedly, \# queries, and \# of relevant papers curated to read.}
\label{tab:my-table}
\end{table*}

One interesting but unplanned observation was that five of the participants were planning to conduct literature reviews for their upcoming paper submissions and expressed interest in using Relatedly after the study had concluded. We saw this as an opportunity to better understand how \textit{Relatedly} performs for real-world tasks over a prolonged period of time.
Therefore, we continued to allowed them access to Relatedly from their computers to conduct their own tasks, and scheduled interviews with them after two weeks to learn about their user experience. The first author then open coded the transcripts of these interviews to identify common themes. All participants were Ph.D. students across three large research universities with an average age of 25.7 years. Two identified as male and three as female. Table~\ref{tab:my-table} shows an overview of their real-world tasks and their engagements with Relatedly.

These scholars expressed their preferences for reading related work sections in Relatedly over reading individual papers to \textit{``learn about a topic''} (P09), and \textit{``verify whether I have covered all the important research threads, papers and perspectives when writing my paper''} (P13)
, \textit{``discover what are the central papers on this topic''} (P15). They also mentioned that it is still helpful to use traditional scholarly search engines to look up specific papers or author, and saw the two approaches as complementary to each other.
P13 summarized their experience by comparing it to their previous literature review process: 
\begin{quote}
``\textit{there's a learning curve to getting used to the shift from reading papers individually one by one to reading paragraphs instead, but once you get used to this paradigm, it's easier to explore the topic this way}.'' (P13)
\end{quote}

Most other benefits mentioned in the post-interviews echoed benefits uncovered in the lab study. Additionally, participants reported using Relatedly with other apps such as: scholarly search engines and PDF readers to read the papers thoroughly; and reference managers like zotero , note-taking apps like notion, documents to attach notes and curate papers for later use.
We see this as a promising signal that \textit{Relatedly} is also able to support real-world tasks over a prolonged period of time, and the fact that participants from Study 1 volunteered to use \textit{Relatedly} after the study under no obligations nor compensation suggested that it provided real-world value to at least the five scholars.


\section{Discussion}
\label{sec:discussion}
This paper illustrates opportunities of leveraging prior scientific effort (i.e., existing related work section paragraphs) to scaffold the users in discovery and synthesis for literature reviews. 
Motivated by a formative study that identified challenges in this approach, we design a novel system, Relatedly, which provides scaffolding features such as auto-generated descriptive paragraph titles, high and low-lighting paragraph text to facilitate reading overlapping and novel information, re-ranking paragraphs to maximize subtopic breath while allowing users to drill-down to specific subtopics. Here, we draw connections between observations from the user study and theory to further contextualize our findings.


Educational psychologists \cite{ainsworth2006deft, rosengrant2007overview} have found that giving students multiple explanations or different representations of the same topic helps them overcome the weaknesses of any particular explanation or representation and better make sense of a complex scientific topic \cite{ainsworth2008educational} and problem solve  \cite{cox1995supporting}. In this context, reading multiple similar paragraphs written by different authors who might frame the topic slightly differently might help users understand and synthesize information topic better. 
Based on the behavior logging, we indeed found that participants explored significantly more paragraphs when using Relatedly than when using the Baseline, which could potentially explain
why they wrote significantly higher quality learning outlines when compared to the Baseline. 

Information foraging theory suggests that people have a tendency to switch between information patches in order to maximize the amount and breadth of information gained during exploration \cite{pirolli1995information}. Relatedly's Overview and Similar Paragraphs Views were designed to facilitate this strategy when exploring multiple related work paragraphs extracted from multiple papers with overlapping and dissimilar information.
During the think-alouds and in the post-task surveys, participants described using the two views to to fluidly alternate between exploring diverse subtopics and exploiting reading details about a specific subtopic, which could potentially explain why they were able to write down significantly more themes/subtopics in their learning outlines when compared to the Baseline.

Finally, participants thought Relatedly was more helpful than the Baseline for literature reviews. Participants in the evaluation study agreed significantly more to the statements: the system helped me ``find relevant research on the topic'', ``understand relationships between terms/concepts'', ``bring together information from multiple sources and points of view'', ``prioritize what to read'' and ``keep track of information gained or read during the literature review process'' compared to the Baseline system. Overall 13 out of 15 participants preferred using Relatedly compared to the Baseline for literature review, and five adopted Relatedly to support their upcoming paper submissions after the study had concluded. Considering this continued usage was volunteered without obligations nor compensations, we see this as a promising indication that Relatedly was able to provide real-world benefits when participants used it to support their own literature review tasks.


We believe that any system that uses algorithmic approaches to help user manage their attentions (recommendations, search, summarization), should be aware of and account for potential model errors misleading users, implicit biases, and echo chamber effects in their designs. When developing Relatedly, we also aimed to mitigate these potential risks. For example, we were careful about the quality of Relatedly's generated section headings and conducted an additional human evaluation in addition to the standard automatic evaluation techniques in NLP (ROUGE). Several of Relatedly's UI features also aimed to combat these risks. For example, the progress bars encourage users to cover more papers and paragraphs instead of feeling satisfied with what they had already explored. The benefits of this increased exploration can help offset possible changes in the distribution of papers explored. Further, while most prior recommender systems help users in finding documents \textit{similar} to what they have already explored, Relatedly, in contrast, actively re-ranks documents and highlight sentences to encourage users to prioritize exploring information \textit{most dissimilar} to what they have already explored.

Relatedly's main insight is that given multiple different texts on a particular topic, it scaffolds the reader's exploration by helping prioritize new dissimilar information and  de-prioritizes redundant information. Relatedly demonstrates this approach using scientific texts and related work sections, however, this approach could be applied in other domains such as policy makers reviewing policy literature for policy briefs, or lawyers researching prior cases to identify patterns, legal precedent and make a case, or voters tracking important issues across multiple politicians' platforms or news articles, or programmers trying to navigate different discussion fora or Jupyter notebooks for solutions for a bug. We envision an augmented reading experience which supports getting a broad overview of different perspective or themes, lets you prioritize what to read and track what you have read so far across information sources on any topic  on the internet. We leave these promising research avenues for future work to explore. 

\subsection{Limitations and Future Work}
Many design decisions were influenced by our focus on literature review of scientific topics by scholars. One assumption we made was that scientists write high quality related work sections in their paper that can provide more benefits to users than looking at individual papers and synthesizing them. For this, we selected papers from leading venues in HCI and NLP to construct our dataset. Future work on this approach that wishes to expand the coverage to all scientific publications would require additional support for finer-grained user control over the sources that the related work paragraphs are extracted from. For example, allowing users to curate a set of venues or authors that they trust. Alternatively, future research could expand on NLP techniques for assessing writing quality \cite{Mesgar2018ANL,taghipour2016neural} to automatically rate the quality of related work sections \cite{teevan_2014} and incorporate them into the ranking algorithm .
Another future direction for further improving Relatedly is to analyze the importance of each reference in the context of the paragraph they were mentioned. For example, NLP techniques such as \cite{Valenzuela2015IdentifyingMC} could be used to estimate the level of influence of each references in a paragraph, so that Relatedly could mitigate the effects of bulk and passing citations \cite{horbach2021meta, kidd1990measuring, horbach2022automated}.


When designing our user study, we also considered citation graph visualization tools as an additional baseline condition. However, literature review tasks can be difficult to study \cite{kules2009designing}, because they can be mentally taxing and time consuming for participants due to their exploratory nature \cite{kules2009designing}. To keep our study realistic, we wanted to keep participants engaged with longer literature review sessions, while keeping the whole procedure under 90-minutes to prevent fatigue.
Therefore, we chose the Baseline condition which simulates the most-popular literature review strategy for most users (i.e., scholarly search engines and reading papers individually). Future work can build on our study by including prior visualization systems as a baseline to compare Relatedly against to help us further understand the costs and benefits of Relatedly.

In the formative interviews, we mostly interviewed PhD students who are junior scholars. Since Relatedly aims to help people jumpstart their lit review process by  broadly overviewing and  finding relevant papers in an unknown topic, we focused on understanding the needs of junior researchers. While three of the participants in the formative study were full-time, post-graduate researchers, we focused on junior researchers for the formative study because they tend to face more challenges and need more support with the literature review process \cite{fitzgerald2017information, ince2018study, davidson2008provenance, tancheva2016day} compared to senior scholars. Senior scholars are more likely to rely on wider social connections to support paper recommendations and have richer adjacent domain knowledge to draw from \cite{davidson2008provenance, fitzgerald2017information}. Existing research systems support senior researchers' literature searches by recommending papers based on social signals such as who they have collaborated or interacted with before \cite{kang2022you, chi23_comlittee}, and papers, authors, institutions, venues of work that they have read or curated \cite{kaur2022feedlens}.

To avoid potential unintended consequences such as plagiarism, Relatedly's design aims to highlight the provenance of ideas and encourage correct referencing practice by prioritizing author information at the top of paragraphs and attaching author information when users copy over references.

One potential obstacle to broader adoption of this approach is licensing and access to scientific documents. Specifically, not all scholarly papers can be freely accessed and searched by anyone on the internet. On the other hand, recent trends in promoting \emph{open science} \cite{mckiernan2016open} and efforts such as the S2ORC dataset \cite{lo-etal-2020-s2orc}, arXiv.org \cite{mckiernan2000arxiv,ginsparg2011arxiv}, and the Open Science Foundation\footnote{Open Science Foundation: \url{https://www.cos.io/products/osf}},  and making older articles accessible using technology such as OCR, GROBID \cite{lopez2009grobid}, VILA \cite{shen2022vila},  point to a promising future where scholars can take fuller advantage of each others prior research effort, enabling new technologies and interactions such as Relatedly. 

Currently, Relatedly is designed for scholarly users. However, an interesting future direction could be supporting lay-people to make exploring scientific information more accessible. For example, if an individual wanted to overview scientific literature on vaccines to determine whether or not to get vaccinated or if they want to overview literature to apply scientific research as a startup product. 
The opportunity here is that seeing different perspectives from authors of different papers describing a research topic and each other's work has the potential of avoiding lay-people overly trusting a single piece of evidence \cite{fischhoff2013sciences}. In this case, a future version of Relatedly could help not only link unfamiliar terminology to definitions \cite{head2021augmenting} or summarize paragraphs in plain language \cite{august2022paper}, but more importantly also surface agreements and disagreements between prior works and their levels of uncertainty while helping users build confidence and trust about their learning could be especially important \cite{fischhoff2013sciences}. 

Knowledge work and literature reviews usually involve exploring multiple topics by issuing multiple queries \cite{fitzgerald2017information, palani2021active}. This is evidenced by the results of the participants who volunteered to use Relatedly for their real-world tasks. While our user evaluation lab study observed how Relatedly helped participants explore information on a single topic and query, an exciting avenue for future work is investigating how users shift their exploration over multiple topics and queries. Recent work, like \cite{palani2021conotate, palani2022interweave}, help people exploring a new domain articulate queries when they lack domain-specific language and well-defined informational goals. We leave it to future work to extend Relatedly's approach to better support exploring scientific literature over multiple queries and topic shifts.




\section{Conclusion}
\label{sec:conslusion}

In this paper we explore a novel approach for supporting literature review workflows---instead of focusing on making sense of individual papers one-by-one to understand a topic, \textit{Relatedly} guided users to explore different subtopic areas using many related work section paragraphs extracted across multiple papers. The idea here is that by leveraging prior scientific efforts of authors conducting literature reviews to write their related work sections, we can improve other researchers' literature review process. To address how paragraphs extracted from different documents might cover both similar and distinct topics, \textit{Relatedly} also provides reading support cues and information gain tracking features to facilitate users in reading many related work paragraphs to cover a broad overview of different topics more efficiently. A comparative user study demonstrated that Relatedly's approach to literature review helped scholars synthesize information on the topic in broader, more coherent and insightful manner. This might have been because Relatedly's reading support features scaffold discovering and interacting with more paragraphs and papers, which helps explore broad multifaceted information spaces. Additionally, participants discovered more domain-specific terms when using Relatedly and preferred using it over the Baseline. We believe the Relatedly approach brings  us one step closer to leveraging information and structure available on the web to support knowledge exploration and synthesis.


\label{sec:acknowledgements}
\begin{acks}
This project is supported by NSF Grant OIA-2033558. The authors would like to thank Daniel S. Weld and Steven P. Dow for the insightful discussions and feedback. We also thank the anonymous reviewers for their constructive feedback. Finally, this work would not have been possible without our pilot test and user study participants. %
\end{acks}
\bibliographystyle{ACM-Reference-Format}
\bibliography{acmart.bib}

\newpage

\appendix
\label{sec:appendix}
 
\begin{table*}[hpt!]
\begin{tabular}{l l}
\hline
Author-Written Titles                                 & Model-Generated Title                                       \\
\hline
Unsupervised Summary Generation & Unsupervised Abstractive Summarization \\
Bezel-initiated Text Entry                   & Text Entry on Smartwatches                        \\
Robots as Social Proxies                     & Designing Social Robots for Social Representation \\
Makers and Makerspaces                       & Makerspaces as Sites of Creativity                \\
Sociocultural Factors and Checklist Efficacy & Cultural Tensions around AI Fairness              \\
Data Table Extraction and Cleaning           & Classification of Web Tables\\                     
Bias in Bilingual Word Embeddings & Bilingual Word Embeddings \\
\hline
\end{tabular}
\caption{Example model-generated headings for paragraphs with long and descriptive author-written titles side-by-side.}
\label{tab:tab_long_heading}
\end{table*}

\begin{table*}[ht!]
\begin{tabular}{ll}
\hline
Author-Written Titles                                 & Model-Generated Title                                       \\
\hline
Related Work   & Machine Translation Optimization        \\
Lucid Dreaming & Lucid Dreaming and Virtual Reality      \\
About Soylent  & Soylent as a Product and Concept        \\
Introduction   & Topic Modeling for Text Segmentation    \\
Definitions    & Delays for Visual Search and Navigation \\
CONCLUDING IMPLICATIONS&  The Moral Economy of Data Management \\
Related Work   & Classic Keyphrase Extraction Systems   \\
\hline
\end{tabular}
\caption{Example model-generated headings for paragraphs with short and generic author-written titles side-by-side.}
\label{tab:tab_short_heading}
\end{table*}

\begin{figure*}[h!]
\centering
\includegraphics[width=\textwidth]{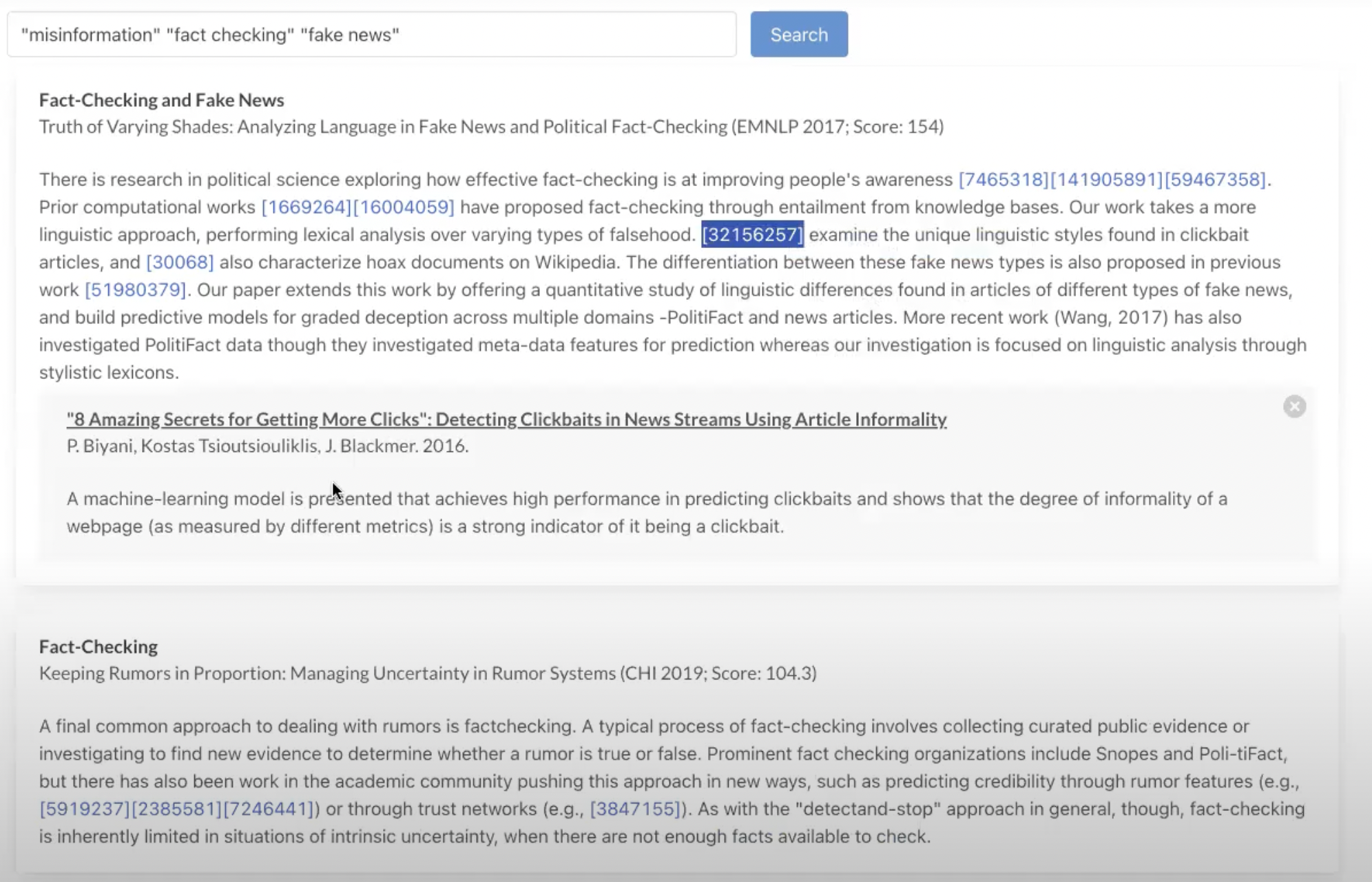}
\caption{During the formative user study, participants were given access to a prototype system where they could search topics and it would return topic-relevant paragraphs from related work sections across multiple paper. They could click on references (hyperlinked corpusIDs) to see paper details (including paper title linked to paper, authors, abstract's TLDR)}
\label{fig:formative}
\end{figure*} 

\begin{table*}[b!]
\begin{tabular}{@{}lllll@{}}
\toprule
 & Relatedly & Baseline & \textit{z} & \textit{p} \\ \midrule
Mental & 14.52 ± 4.77 & 16.43 ± 4.35 & -0.94 & 0.36 \\
Physical & 13.33 ± 4.67 & 15.24 ± 4.29 & -0.86 & 0.33 \\
Temporal & 12.39 ± 3.45 & 14.04 ± 2.49 & -0.83 & 0.37 \\
Performance & 18.37 ± 4.34 & 12.00 ± 2.79 & -0.67 & 0.49 \\
Effort & 12.86 ± 4.79 & 14.05 ± 4.22 & 0.24 & 0.80 \\
Frustration & 6.83 ± 4.30 & 9.97 ± 6.34 & 1.59 & 0.11 \\ \bottomrule
\end{tabular}
\caption{There were no significant differences in task workloads when using Relatedly vs Baseline suggesting improved performance with similar precieved workload. We report the mean NASA-TLX scores with standard deviation as uncertainty and results from Wilcoxon Signed-Rank tests, with z and p values. The range of possible values is 1-20. }
\label{tab:nasatlx}
\end{table*}

\end{document}